\def\to{\hbox{$\,$--$\,$}}
\def\minus{\hbox{$\,$--$\,$}}
\def\muspc{\hskip 0.15 em}

\def\mag{\hbox{$\;.\!\!\!^m$}}
\documentstyle[11pt,paspconf]{article}
\input psfig
\begin{document}

\title{\hfill Are there carbon stars in the Bulge ?\hskip 1.5cm\null}
\author{\vskip -1.5cm
\vbox{\psfig{file=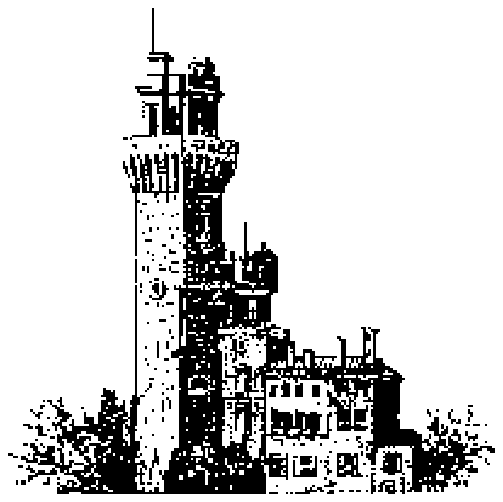,height=10.0cm,width=10.0cm}}
\vskip-9.0cm\null
\hfill
Yuen Keong Ng\hfill\null\vskip 0.3cm
}
\affil{Padova Observatory, 
Vicolo dell'Osservatorio 5, I-35122 Padua, Italy}

\begin{abstract}
The bulge carbon stars have been a mystery since their discovery
by Azzopardi et~al.\ (1991), because they are about 
2\mag5 too faint to be regarded as genuine AGB stars, if located
inside the metal-rich bulge (\mbox{m{\to}M\muspc=\muspc14\mag5}).
Part of the mystery can be solved if these carbon stars 
are related to the Sagittarius dwarf galaxy (SDG; 
\mbox{m{\to}M\muspc$\simeq$\muspc17\mag0}). 
They are in that case not old and metal-rich,
but young, $\sim$0.1~Gyr, with SMC-like metallicity (Ng 1998).
\hfill\break
The $\sigma_{RV}\!=\!113\pm14$~km s$^{-1}$
(Tyson{\muspc\&\muspc}Rich~1991) 
radial velocity dispersion
of the stars appears to be consistent with 
bulge membership. On the other hand, a similar velocity dispersion
could be the result from an induced star formation event when the
SDG crosses the galactic midplane.
It is suggested that the
carbon stars are tracers of such an event and that they
therefore are located at distances related to the SDG. However,
the majority of the carbon stars are not member of the SDG,
nor are they similar to the C-stars which are member of the SDG.
\hfill\break
The radial velocities can be used to determine a possible membership to the
SDG. However,
they do not give information about the distance of the stars.
In particular, if the stars are located
at a distance comparable to the SDG.
This implies that only the period-luminosity relation 
(Groenewegen{\muspc\&\muspc}Whitelock 1996) can be used to distinguish 
unambiguously if the carbon stars are located at bulge-like
or SDG-like distances. 
Thus far only carbon stars with reliable periods have been identified 
at a SDG related distance 
(Ng{\muspc\&\muspc}Schultheis 1997, Whitelock 1998).
\end{abstract}


\section{Introducing the mystery}
After a long search 34 carbon stars found by 
Azzopardi et~al.\ (1991 and references cited therein --
ALRW91) in the direction
of the Galactic Centre. 
If located in the metal-rich Bulge they are 
about \mbox{2\mag5} fainter than the 
carbon stars from the Magellanic
Clouds or the dwarf spheroidal galaxies
(Lequeux 1990, Tyson{\muspc\&\muspc}Rich 1991 -- TR91, 
Westerlund et~al.\ 1991). \hfill\break
The Bulge is older than the maximum age of 
\mbox{$\sim$\muspc4~Gyr} 
for a carbon star to be regarded as a genuine TP-AGB star
(see Ng 1997 \& 1998 and references cited
therein). The ALRW91 C-stars are therefore thought 
to be located along the RGB as a result
of binary evolution.
Figure~1 indicates that for the latter case the stars do not 
reach the RGB tip.
Moreover, the origin of this truncation has
not been explained.
\hfill\break
The questions raised and answered partly to provide a solution to
this standing mystery are: 
are these stars really located in the Bulge~? 
Is there an alternative~? If so:
where are they~? and what are the
implications~?

\begin{figure}
\setbox1=\vbox{\hsize=4.0cm\psfig{file=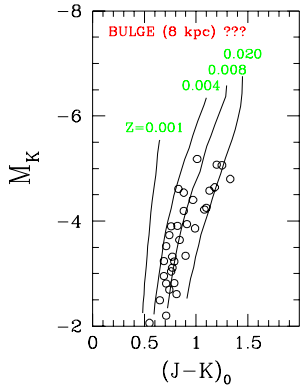,height=5.1cm,width=3.9cm}}
\setbox2=\vbox{\hsize=8.5cm\caption{%
The \mbox{(M$_K$,\/J--K)$_0$} CMD 
of the RGB for 10~Gyr isochrones for
the metallicities \mbox{Z\muspc=\muspc0.001}, 0.004, 0.008, and 0.020
(Bertelli et~al.\ 1994).
The absolute magnitudes of the ALRW91 C-stars
are determined under the assumption that they are located in the
Bulge at a distance of 8~kpc.
Note that the ALRW91 C-stars do not reach the RGB tip
and no explanation has been provided thus far
for this behaviour. 
}}
\centerline{\copy1\quad\copy2}
\end{figure}

\section{Relation to the Sagittarius dwarf galaxy ?}
The serendipitous identification of the 
Sagittarius dwarf galaxy (SDG) was made by 
Ibata et~al.\ (1994, 1995).
Accurate distance determinations from RR Lyrae stars belonging to this galaxy
range from 22.0\to27.3~kpc (Ng~1997 and references cited therein).
The $\sim$2\mag5 difference of the distance modulus 
between the dwarf galaxy 
(\mbox{m{\to}M\muspc=\muspc17\mag0})
and the Galactic Centre at 8~kpc
(\mbox{m{\to}M\muspc=\muspc14\mag5})
lead Ng{\muspc\&\muspc}Schultheis (1997) 
to suggest that the ALRW91 C-stars
could actually be located at the distance of the 
dwarf galaxy. 
Note that this does not necessarily imply a SDG membership.
Its presence
was unknown at the time when the C-stars were identified
and a different location 
could solve the standing question about the origin 
of the `bulge' carbon stars.

\begin{figure}
\centerline{\vbox{\psfig{file=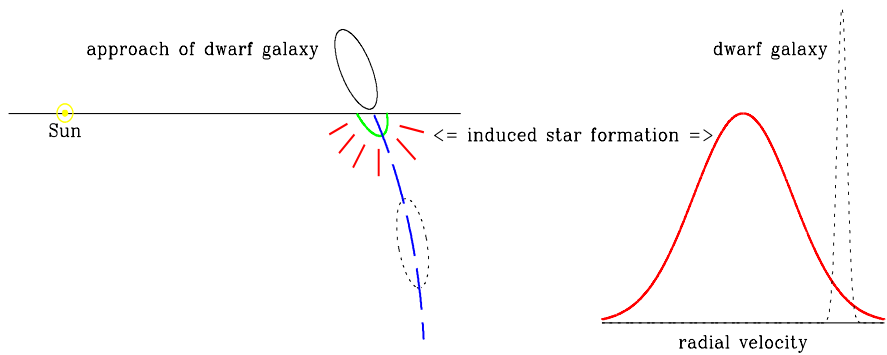,height=4.6cm,width=11.8cm}}}
\caption{Schematic view of the broad, bulge like, radial 
velocity distribution for the material from an induced star 
formation event, caused by the crossing from a dwarf galaxy 
through the Galactic mid-plane.
Note that the majority of the new material will not be moving 
along the dwarf galaxy orbit. The dotted ellipse and the 
dotted line indicate respectively the 
current position and radial velocity (not on scale) for the dwarf galaxy}
\end{figure}

\section{Not so metal-rich after all~!}
Ng (1997 \& 1998) analysed this possibility by 
assuming a relation of the ALRW91 C-stars 
with the SDG. Once more it is stressed that 
{\it related $\ne$ membership},
except for a small fraction of the C-stars.
Ng demonstrated that the photometric sequence of the ALRW91 C-stars 
is in that case not exceptional, but comparable with the 
sequence found for the SMC. 
In addition, the ALRW91 carbon stars are not 
metal-rich as previously thought. 
The metallicity was initially estimated to
\mbox{Z\muspc$\simeq$\muspc0.008}
(Ng 1997), but a revision has been made to
\mbox{Z\muspc$\simeq$\muspc0.004} 
(Ng 1998). Ng further argued that the stars have 
an age \mbox{$\sim$\muspc0.1~Gyr}.

\section{Elusive radial velocities}
The radial velocities from TR91 for
the carbon stars appear to be consistent with a Bulge membership.
However, an alternative explanation for the large velocity dispersion
is available: induced star formation  
from the material residing in either the SDG and/or the impact spot 
in the Galactic disc when the SDG respectively 
approaches and crosses the galactic plane (Ng 1997 \& 1998).
Figure~1 displays a schematic view of the process, where 
the SDG pushes material out of the Galactic 
disc\footnote{Numerical calculations of the structure of the 
H~I disc by Ibata\mbox{\muspc\&\muspc}Razoumov (1998)
indicate that just after impact (their Fig.~2b) material 
is scattered in all directions above and below the Galactic plane. 
This should result in a broad radial velocity of the 
material scattered after the crossing of the Galactic plane.
Their Fig.~4 further indicates that part of the gaseous material
with log~T in the range 2\to4 should have been converted into stars;
some of them possibly comparable to the ALRW91 C-stars.
}. 
Only a small fraction of the newly 
formed stars is dragged along the orbit, while another fraction
is moving away from the SDG.
Note that another fraction of these young stars is moving 
towards us. This implies that the distribution
of radial velocities is considerably broader than expected 
from a SDG membership alone, i.e.
\mbox{$\sigma_{\rm RV}$\muspc$>>$\muspc11.4\muspc$\pm$\muspc0.7~km~s$^{-1}$}.
The TR91 velocity dispersion of 
\mbox{$\sigma_{\rm RV}$\muspc=\muspc113\muspc$\pm$\muspc14~km~s$^{-1}$}  
for the ALRW91 C-stars may have been interpreted mistakingly 
as an indication for a Galactic Bulge 
membership\footnote{The uncertainties in the individual velocity
values may be as large as \mbox{90~km~s$^{-1}$}. However,
if the uncertainties follow a random pattern the mean velocity
and its dispersion will not change drastically if the
radial velocity values are improved. 
}. 
Alternatively 
the large velocity dispersion can also be a signature from 
an induced star formation event.

\section{Gateway: the Galactic disc and the Sagittarius dwarf galaxy} 
The impact position at the galactic mid-plane 
is obtained from the unweighted least-squares
fit through the distances determined from
the SDG RR~Lyrae stars
as a function of the galactic latitude (Ng 1998)
and yields 22.8~kpc.
Adopting a Solar galacto-centric distance of 
\mbox{R$_0$\muspc=\muspc8.0~kpc} 
implies that the crossing of the SDG 
occurred at 14.8~kpc distance from
the Galactic centre.
\par
To determine the disc metallicity at impact position one 
has to take the radial dependence of [Fe/H] into account.
Carraro et~al.\ (1998) determined from open clusters 
the radial metallicity gradient in different age ranges. 
The disc metallicity at the crossing position is
\mbox{[Fe/H]\muspc=\muspc\to0.61\muspc$\pm$\muspc0.20}
(\mbox{Z\muspc$\simeq$\muspc0.0050})
or \mbox{[Fe/H]\muspc=\muspc\to0.73\muspc$\pm$\muspc0.10} 
(\mbox{Z\muspc$\simeq$\muspc0.0035})
with respectively the present day and the
average radial metallicity gradient. 
\hfill\break
Irrespective of the present day or average metallicity 
gradient it is estimated that the disc metallicity at the 
crossing position is 
\mbox{Z\muspc$\simeq$\muspc0.0045\muspc$\pm$\muspc0.0010}.
Within the uncertainties this is in agreement with  
\mbox{Z\muspc$\simeq$\muspc0.004}.
\par
Considering that an induced star formation event can explain the 
large velocity dispersion for the ALRW C-stars
and taking into account the 
similarity between the metallicity from a photometric estimate and 
the disc impact position, it is suggested that the C-stars
do not originate from the SDG itself, but are likely formed 
from material originating from our own Galactic disc.
\par
The average (J{\minus}K)$_{0,saao}$ colour of the carbon stars 
which are member of the dwarf galaxy is 1\mag40
(Whitelock et~al. 1996), while the 
average (J{\minus}K)$_{0,eso}$ colour of the ALRW91 C-stars 
is 0\mag85. This immediately shows 
the marked difference between two groups of carbon stars
found in the same direction at a comparable distance.
The redder SDG C-stars are likely older and metal-richer 
than the ALRW91 C-stars, because
the redder SDG C-stars should be associated with a SDG population
comparable or younger than \mbox{$\sim$\muspc4~Gyr}.
The youngest population identified for the SDG has an age 
of about 4~Gyr and a metallicity \mbox{Z\muspc$\simeq$\muspc0.008}
(Mighell et~al.\ 1998). 

\section{Why are they still around Sagittarius dwarf galaxy?}
The average radial velocity of \mbox{\minus44~km~s$^{-1}$} (TR91)
provides a first indication of the distance traveled  
by the ALRW91 C-stars.
Considering a smaller velocity in the past
the distance traveled in 0.1~Gyr is 
\mbox{$\sim$\muspc2.2\to3.1~kpc}. 
The traveled distance is within the \mbox{10\%\to15\%}
uncertainty of the 22.8~kpc distance to the impact position.
\par
The velocity dispersion of the C-stars provides further an indication
about the average separation. Taking into
account that the dispersion is smaller,
due to a combination of turbulence and a lower velocities in the past,
the separation is estimated to \mbox{$\sim$5.6\to8.0~kpc}.
For comparison:
the distance between 2 of the 4~globular clusters associated 
with SDG is about 10~kpc,
i.e. Terzan 8 at 21.1~kpc and Arp~2 at 31.0~kpc 
(Da Costa{\muspc\&\muspc}Armandroff 1995),
while the distance between the two carbon stars with well determined
periods is about 5 kpc, i.e. 
21.9~kpc for a carbon semiregular variable 
(Ng\mbox{\muspc\&\muspc}Schultheis 1997), 
and 26.7~kpc for a carbon Mira (Whitelock 1998).
\par
Both the average distance traveled by the ALRW91 C-stars 
and the separation between them is such that most of the
stars can still be associated with the Sagittarius dwarf galaxy.

\section{Still mysterious ???} 
The general accepted models of the SDG orbit indicate 
(see Fig.~11 Ibata et~al.\ 1997 and references cited therein), 
together with the proper motion reported by 
Irwin et~al.\ (1996), that
the SDG is still moving towards the galactic mid-plane.
On the other hand,
the ALRW91 C-stars appears to form the factual evidence that the 
SDG has already crossed the galactic mid-plane.
In addition, the orbit 
at low galactic latitudes (Fig.~4, Ng 1997) is not
consistent with the position obtained from RR~Lyrae stars,
which might partly due to uncertainties in the extinction.
A study of the Galactic globular cluster Palomar~5 
(Scholz et~al.\ 1998) further indicates that the cluster 
is moving in the opposite direction with respect to the
Ibata et~al.\ orbital motion of the SDG. 
\hfill\break
Instead of looking for alternative explanations for the 
apparent contradiction, one might consider a re-analysis of the orbit
combined with an independent verification of the proper motion.
A revised orbit and proper motion might show that the  
globular cluster Palomar~5 is tidally stripped from the
SDG (Lin 1996) and also associate the 
Smith high velocity Cloud with the SDG (Bland-Hawthorn et~al.\ 1998).
\par
The models thus far consider only the gravitational interaction between
the Galaxy and the SDG (Ibata\mbox{\muspc\&\muspc}Lewis 1998,
Nair\mbox{\muspc\&\muspc}Miralda-Escud\'e 1998), 
but do not take into account
an encounter, induced, star formation event.
Improved models should therefore be 
employed to trace the stars
from such an event.
In addition, a detailed proper motion
study and a renewed determination of the radial velocities
of the ALRW91 C-stars are still required to explain 
and better understand the presence of the carbon stars. 
\par

\section{Conclusion} 
The `Bulge' (ALRW91) carbon stars are located at a distance comparable 
to the Sagittarius dwarf galaxy. However, the majority of these stars
(\mbox{Z\muspc$\simeq$\muspc0.004}, \mbox{t$\simeq$0.1~Gyr})
are not member of the Sagittarius dwarf galaxy nor are they 
comparable with the carbon stars found in the Sagittarius dwarf galaxy
(\mbox{Z\muspc$\simeq$\muspc0.008}, \mbox{t$\simeq$4~Gyr}; 
see for details Ng 1998).
\par
The radial velocities cannot be used reliably as an argument 
in favour of Bulge membership, due to a possible confusion with the
velocity dispersion of the stars resulting from a crossing induced
star formation event which scatters new stars in all 
directions\footnote{Bonus: the microlensing events between the
stars from the bulge/bar and the induced semi-sphere of young stars,
due to the crossing of the SDG through the galactic midplane,
could well prove to contribute significantly to the surprisingly high
event rate found towards the 
Galactic Centre (Paczy\'nski 1996 and references cited therein).}.
\hfill\break
If there are carbon stars in the Bulge then 
the carbon star period-luminosity relation 
(Groenewegen\mbox{\muspc\&\muspc}Whitelock 1996) 
should be employed to determine their distances.
Up to date no carbon semiregular or Mira variables have been
reported with a period-luminosity which unambiguously associates them 
with the Galactic Bulge.

\acknowledgments
Ng acknowledges a travel grant received from the IAU.
The research is supported by grant ERBFMRX-CT96-0086 from
the TMR network {\it`Formation and evolution of galaxies'}\/
of the European Community.

\vfill
\end{document}